\documentclass{elsart3p}
\usepackage{amssymb}
\usepackage{graphicx}

\begin{document}

\begin{frontmatter}
\title{Anomalous bond stretching phonons as a probe of charge fluctuations in perovskites}
\author[a,c]{S. Cojocaru \corauthref{cor}}
\corauth[cor]{Corresponding author.}
\ead{cojocaru@sa.infn.it}
\author[a]{R. Citro}
\ead{citro@sa.infn.it}
\author[a,b]{M. Marinaro}
\ead{marinaro@sa.infn.it}
\address[a]{Dipartimento di Fisica "E. R. Caianiello",
Universit\'{a} degli Studi di Salerno and CNISM, Unit\'{a} di
ricerca di Salerno, Via S. Allende, 84081 Baronissi (SA), Italy}
\address[b]{I.I.A.S.S., Via G. Pellegrino, n. 19 84019 Vietri sul Mare (SA), Italy}
\address[c]{Institute of Applied Physics, Chisinau 2028, Moldova}

\begin{abstract}
Important information on momentum resolved low energy charge
response can be extracted from anomalous properties of bond
stretching in plane phonons observed in inelastic neutron and
X-ray scattering in cuprates and some other perovskites. We
discuss a semiphenomenological model based on coupling of phonons
to a single charge mode. The phonon dispersion and linewidth allow
to locate the energy of the charge excitation in the mid infrared
part of the spectrum and to determine some of its characteristics.
New experiments on oxygen isotope substitution could allow to
achieve a more detailed description. Corresponding relations
following from the model can be used for the interpretation of
experiments and as test of the model.
\end{abstract}
\end{frontmatter}

\section{Introduction}

A direct coupling of the bond stretching (BS) inplane ($CuO_{2}$)
phonon modes to density fluctuations of strongly correlated
electrons has been
discussed as a possible reason of their anomalous properties \cite{Khal},%
\cite{Gun}. Doping dependent anomalies of these high energy
($70-90$ meV) phonon branches are observed in a number of
perovskite materials like cuprates, bismuthates, manganites,
nickelates \cite{P2}-\cite{Reznik}. In particular, presence of a
strong (more than $20\%$) softening of the BS phonon dispersion
towards the Brillouin zone boundary in the $\left( 100\right) $
and $\left( 010\right) $ directions contrasts a very week
dispersion in the diagonal direction $\left( 110\right) .$ This is
paralleled by the phonon linewidth that reaches unusually large values too, $%
10$ meV, in the $\left( 100\right) $ and $\left( 010\right) $. In
some materials softening can also be non-monotonous and one
observes a minimum at intermediate momenta, e.g. $q\sim 0.3$
r.l.u. A number of theories and models have been developed to
explain the properties of BS phonons. To mention a few possible
scenarios, such as incommensurate charge density wave instability
or stripes, acoustic and optical plasmons or unusual dielectric
properties (overscreening, when the dielectric constant becomes
negative at
some momenta) as well as calculations based on ab-initio LDA, e.g. \cite%
{Falter}, or strong electron correlations, e.g. \cite{Kulic}.
Different scenarios refer to different energy scale of the charge
excitations coupled to BS phonons. In most cases its direct
experimental measurement is not available, because it requires a
sufficient momentum resolution at relatively low energies. Most of
the present knowledge comes from optical experiments, i.e. small
momentum $q,$ where the coupling to the BS phonon mode is
vanishing. In this context it seems interesting to consider the
phonon itself as a probe of the charge excitation. We employ a
semiphenomenological model where the charge response coupled to
the phonon is parametrized in a simple form. To determine all
three parameters some additional measurements of the BS mode
properties are required. We propose to consider dependence on
oxygen isotope substitution of phonon dispersion and linewidth.
This would allow not only to determine the parameters of the model
but also to test its validity, since the number of equations is
larger than required by selfconsistency.

\section{Model and results}

The Dyson equation for the renormalized phonon propagator can be
written in
the form%
\begin{equation}
\omega_{0}D^{-1}\left( q,\omega\right)
=\omega^{2}-\omega_{0}^{2}\left( 1+\alpha\sin^{2}\left(
q_{x}/2\right) P\left( q,\omega\right) \right) , \label{Phonon}
\end{equation}
where the structure factor corresponds to $\left( 100\right) $,
$\omega_{0}$
is the bare frequency, $\alpha$ is the coupling constant \cite{Khal} and $%
P\left( q,\omega\right) $ the charge susceptibility. The latter is
taken in
the Lorentzian form%
\begin{equation}
P\left( q,\omega\right) =\frac{\eta_{q}}{\omega^{2}-\Omega_{q}^{2}+i%
\Gamma_{q}\omega},   \label{mode}
\end{equation}
with unknown dispersion $\Omega_{q},$ linewidth $\Gamma_{q}$ and
oscillator strength $\eta_{q}.$ The model (\ref{Phonon}),
(\ref{mode}) defines the
equation for the phonon frequency%
\begin{equation}
\omega_{q}^{2}=\omega_{0}^{2}\left(
1+\beta_{q}\frac{\sin^{2}\left( q_{x}/2\right) \left(
\omega_{q}^{2}-\Omega_{q}^{2}\right) }{\left(
\omega_{q}^{2}-\Omega_{q}^{2}\right) ^{2}+\left( \Gamma_{q}\omega
_{q}\right) ^{2}}\right) ,   \label{q1}
\end{equation}
and linewidth%
\begin{equation}
\gamma_{q}=\Gamma_{q}\frac{\omega_{0}^{2}-\omega_{q}^{2}}{%
\Omega_{q}^{2}-\omega_{q}^{2}},   \label{Damping}
\end{equation}
where $\beta_{q}=\alpha\eta_{q}$ is the effective coupling
constant. Analysis of above equations shows that according to
observed features of the BS softening, dispersion $\Omega_{q}$
along $\left( 100\right) $ and $\left( 010\right) $ should be much
smaller than in diagonal direction, that agrees with results based
on $t-J$ model \cite{Khal}. We then estimate the values of
$\Gamma_{q}$ by substituting known data on $\omega_{q}$ and
$\gamma_{q}$ into (\ref{Damping}) and changing the values of
$\Omega_{q}.$ After these results are confronted with data on
optical absorption, it follows that the
relevant charge excitation should be located in the mid infrared region, $%
0.2-0.6$ eV. This is further corroborated by the analysis of its
doping dependence in some cuprates \cite{Hanke} as compared to
softening of the BS phonons.

Additional information can be extracted by considering the effect
of the isotope mass shift $dm=m\left( O^{18}\right) -m\left(
O^{16}\right) $ on dispersion $d\omega _{q}=\omega _{q}\left(
O^{18}\right) -\omega _{q}\left(
O^{16}\right) $ and linewidth $d\gamma _{q}$. In particular, the model yields%
\begin{equation}
d\omega _{q}/d\omega _{0}\simeq \sqrt{1-\beta _{q}\sin ^{2}\left(
q_{x}/2\right) /\Omega _{q}^{2}},  \label{e1}
\end{equation}%
\begin{equation}
d\gamma _{q}/d\omega _{0}\simeq 2\gamma _{q}/\omega _{0}.
\label{e2}
\end{equation}%
in the lowest order of the $\omega _{0}/\Omega _{q}$ expansion. We
further
define the momentum dependent isotope coefficient $\alpha _{q}\equiv $ $%
-\left( md\omega _{q}\ /\ \omega _{q}dm\right) $ that allows to
introduce a new relation to the parameters of the charge
excitation derived in the same
approximation%
\begin{equation}
\alpha _{q}-\alpha _{0}=\alpha _{0}\beta _{q}\left( \Gamma
_{q}^{2}/\Omega _{q}^{2}-1\right) \sin ^{2}\left( q_{x}/2\right)
\omega _{0}^{2}/\Omega _{q}^{4}.  \label{e3}
\end{equation}%
If $\omega _{0}/\Omega _{q}$ is not small, i.e. dynamic effects
are important, the full expressions should be used (not given
here).

\begin{figure}[tbph]
\centering\includegraphics*[width=0.9\linewidth]{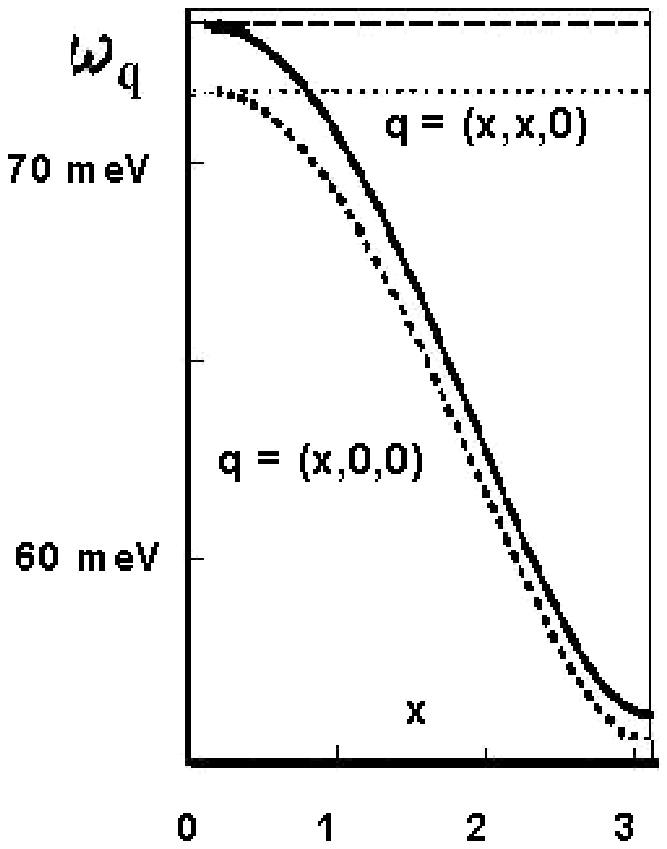}
\caption{The expected isotope dependence of phonon dispersion
along the two directions in the Brillouin zone.}
\end{figure}

Eqs. (\ref{q1}-\ref{e3}) contain the relation between
experimentally measurable characteristics of the phonon spectrum
and the parameters of the model describing the charge excitation.
On the basis of available data the model predicts that the energy
of charge excitation responsible for the renormalization of the BS
phonon mode is located in the mid infrared region. Its energy is
lowered with hole doping and respectively its effect on the BS
phonon increases. To be noted that the sign of the differential
isotope coefficient (\ref{e3}) depends on the ratio of $\Omega
_{q}$ to the linewidth. We expect that the fingerprint of the
broad MIR structure will show up in a positive value of this
coefficient. Another observation is that phonon linewidth could be
a more sensitive probe of the isotope effect than phonon
dispersion as seen from (\ref{e1}) and (\ref{e2}).


\begin{thebibliography}{99}
\bibitem{Khal} P. Horsch and G. Khaliullin, Physica B, \textbf{359-361}, 620
(2005); D.N. Aristov and G. Khaliullin cond-mat/0603161 (and
references therein).

\bibitem{Gun} O. Rosch and O. Gunnarsson, Phys. Rev. Lett. \textbf{93},
237001 (2004).

\bibitem{P2} L. Pintschovius, Phys. Stat. Sol. B \textbf{242}, 30 (2005).

\bibitem{Fukuda} T. Fukuda et al., Phys. Rev. B\textbf{\ 71}, 060501 (2005).

\bibitem{Braden1} M. Braden et al., Physica C, \textbf{378-381}, 89 (2002).

\bibitem{Tranquada} J.M. Tranquada et al., Phys. Rev. Lett. \textbf{88},
075505 (2002).

\bibitem{Reznik} D. Reznik et al., Nature \textbf{440}, 1170 (2006).

\bibitem{Falter} C. Falter, Phys. Stat. Sol. B \textbf{242}, 78 (2005).

\bibitem{Kulic} M. L. Kuli\'{c} and O. V. Dolgov, Phys. Rev.  \textbf{71},
092505 (2005).

\bibitem{Hanke} F. Hanke and M. Azzouz, M. J. Cond. Matt., \textbf{6}, 1
(2005); http://www.fsr.ac.ma/MJCM/vol6-art01.pdf.

\end{thebibliography}
\end{document}